# Controlling the charge environment of single quantum dots in a photonic-crystal cavity

***N. Chauvin*[1], *C. Zinoni*[3], *M. Francardi*[2], *A. Gerardino*[2], *L. Balet*[1,3], *B. Alloing*[3], *L.H. Li*[3], *and A. Fiore*[1]**

*1. COBRA Research Institute, Eindhoven University of Technology, PO Box 513, 5600MB Eindhoven, The Netherlands*

*2. Institute for Photonics and Nanotechnologies-CNR, via Cineto Romano 42, 00156 Roma, Italy*

*3. Institute of Photonics and Quantum Electronics, Ecole Polytechnique Fédérale de Lausanne, CH-1015 Lausanne, Switzerland*

## Abstract

We demonstrate that the presence of charge around a semiconductor quantum dot (QD) strongly affects its optical properties and produces non-resonant coupling to the modes of a microcavity. We first show that, besides (multi)exciton lines, a QD generates a spectrally broad emission which efficiently couples to cavity modes. Its temporal dynamics shows that it is related to the Coulomb interaction between the QD (multi)excitons and carriers in the adjacent wetting layer. This mechanism can be suppressed by the application of an electric field, making the QD closer to an ideal two-level system.

## Main Text

The study of single quantum dots (QDs) embedded inside photonic crystal (PhC) cavities or micropillars has been the subject of an intense interest, for both fundamental science and applications [1]. Indeed, the discrete energy structure in a QD makes it a semiconductor equivalent of an atomic system, and allows solid-state implementations of cavity quantum electrodynamic experiments, with applications to single-photon sources, entanglement generation and quantum computing. However, the micro-photoluminescence experiments performed on QDs coupled to optical microcavities groups have revealed a phenomenon contradicting the ideal atom-in-a-cavity model: the cavity mode emission was observed despite the lack of an excitonic transition in resonance with the mode [2-11]. This behaviour contradicts the two-level system model and represents a problem for the application of QDs to solid-state quantum information processing. Indeed, the cavity emission was observed to provide a classical photon statistics [2], which for example reduces the purity of QD single-photon sources. So far, several physical mechanisms have been evoked to explain this observation: dephasing processes [6,7], a continuum in the hole states [4], a continuum due to a mixing between s and p states [3]. Experimental results have provided evidence of the role of dephasing in nonresonant coupling to cavity modes in the case of small detuning (1 to 3 meV) [5,9-11]. In this paper, we show the existence of a second process which provides nonresonant coupling at much larger detunings (up to 10 meV). It originates from the Coulomb interaction between the carriers in the wetting layer and the multiexcitons in the QD, which generates a spectrally-wide emission coupled to the mode. We support this interpretation by investigating the dynamics of the QD-cavity system. Additionally, we show that the charge environment around the QD can be controlled by the application of an electric field. The selective removal of the nonresonant cavity coupling by the field confirms its physical origin and allows us to approach the ideal situation of an atomic-like emitter in a cavity.

We investigate InAs self-assembled QDs in photonic-crystal (PhC) cavities, although we expect that our conclusions also apply to other types of cavities, such as micropillars and microdisks. A typical low-temperature (5K) emission spectrum from a L3 cavity (3 missing holes [12]), as measured in a microphotoluminescence set-up, is shown in Fig. 1a). The sample, grown by molecular beam epitaxy, consists of a 320-nm-thick GaAs membrane on top of a 1.5 μm $Al_{0.7}Ga_{0.3}As$ sacrificial layer. A single layer of low density (5–7 dots/μm$^2$) self-assembled InAs QDs emitting at 1.3 μm at low temperature is embedded in the middle of the membrane (Ref 12).

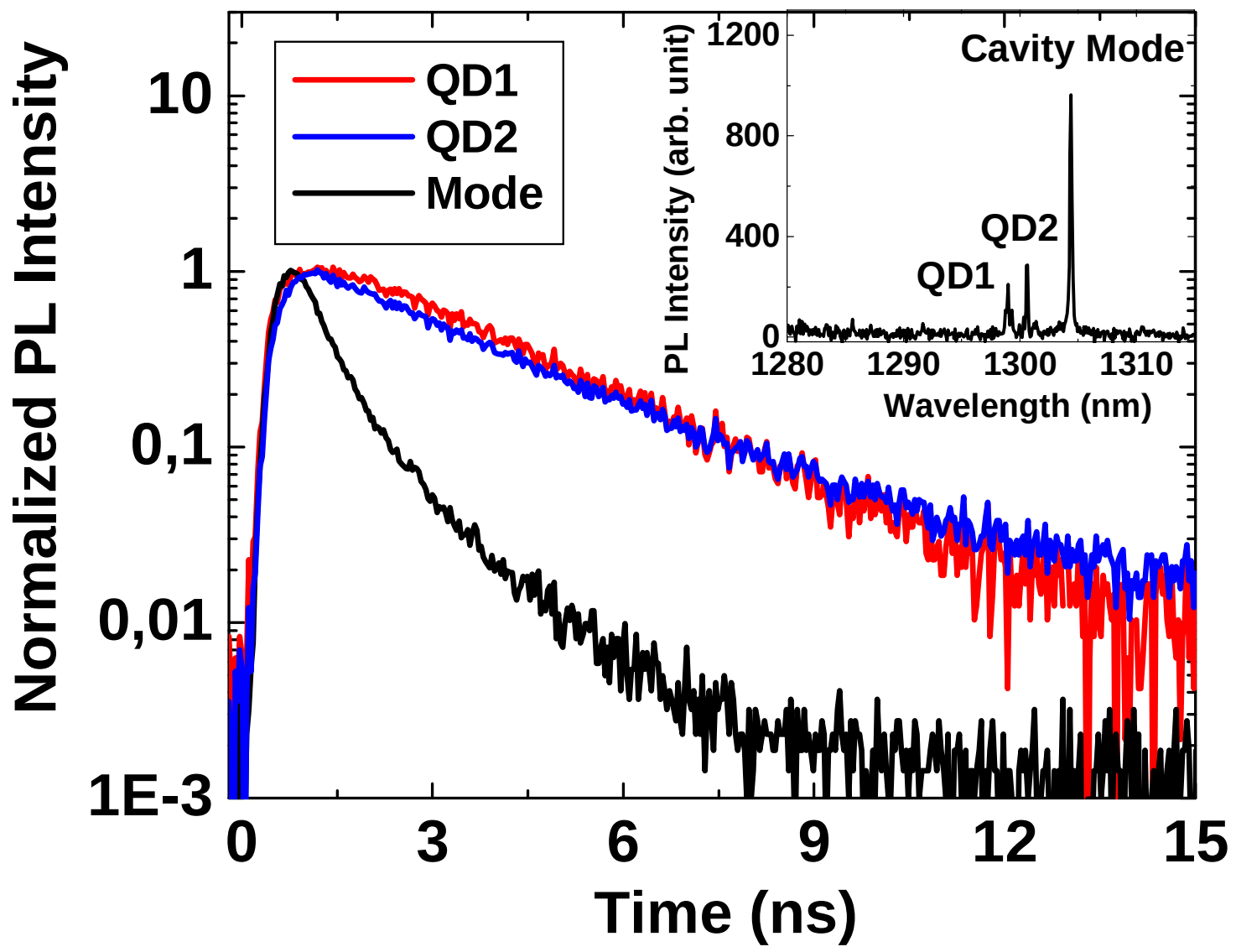

(Color online) Figure 1: Time resolved experiments of the cavity mode and off resonance QDs. Inset: spectrum of a L3 PhC cavity, under pulsed excitation (λ= 750 nm, average pump power $P_{av}$=0.4μW in a 2 μm-diameter spot), showing the cavity mode emission and two off-resonance QDs.

A strong cavity mode emission with a quality factor of 11,500 and two sharp lines associated to single QDs are observed (inset of Fig. 1). The photoluminescence (PL) decay was measured by time-correlated fluorescence spectroscopy using a gain-switched 750 nm pump laser and a superconducting single-photon detector [13], providing a combined temporal resolution of 150 ps (Fig. 1). The excitonic line QD1 (QD2) has a monoexponential decay with a lifetime of 2.6 ns (3.1 ns), lower than the intrinsic radiative time of 1.1 ns [14], due to the low available optical density of states in the off-resonant cavity. In contrast, the cavity mode emission has a biexponential decay with a fast lifetime of 0.4 ns and a slow lifetime of 1.3 ns. 85% of the cavity mode emission comes from this fast decay which is clearly distinct from the decay of the excitonic lines QD1 and QD2. This behavior, typically observed in our samples for detunings up to 10 nm, and already reported in Ref. 12, shows that in these structures the cavity mode emission cannot be due to the homogeneous broadening of the excitonic lines. Indeed, in the latter case both lines would mirror the decay of a single emitter resulting in the same dynamics [10].

We instead propose that the off-resonant cavity peak is pumped by a spectrally-wide emission (to be referred to as “background”), due Coulomb interactions between carriers in the wetting layer and multiexcitons in the QD. Such emission, also previously reported [15-18], is strongly enhanced by the cavity coupling, as also confirmed by the fast lifetime observed in Fig. 1(b). In order to gain more insight in the carrier dynamics within a QD, we have investigated the PL decay of a QD in the absence of a cavity. To this aim a layer of low-density QDs was grown in a lambda/2 planar cavity [19]. Small (1 $\mu m^2$) apertures in a gold mask were processed to isolate a single QD. Figure 2(a) shows the time-integrated spectrum of a single QD under a $P_{av}$= 6μW excitation at 6K. Single lines associated to neutral or charged multiexcitonic states are clearly observed together with a broad background emission, which extends over >10 nm, depending on the excitation density. Figure 2(b) shows a streak-camera like image of the PL decay at different wavelengths, obtained by scanning a tunable fiber bandpass filter (full width at half maximum of 0.3 nm) and measuring the decay at each wavelength by time-correlated fluorescence spectroscopy. The use of a high-sensitivity SSPD was essential to measure the dynamics of the low-level background signal.

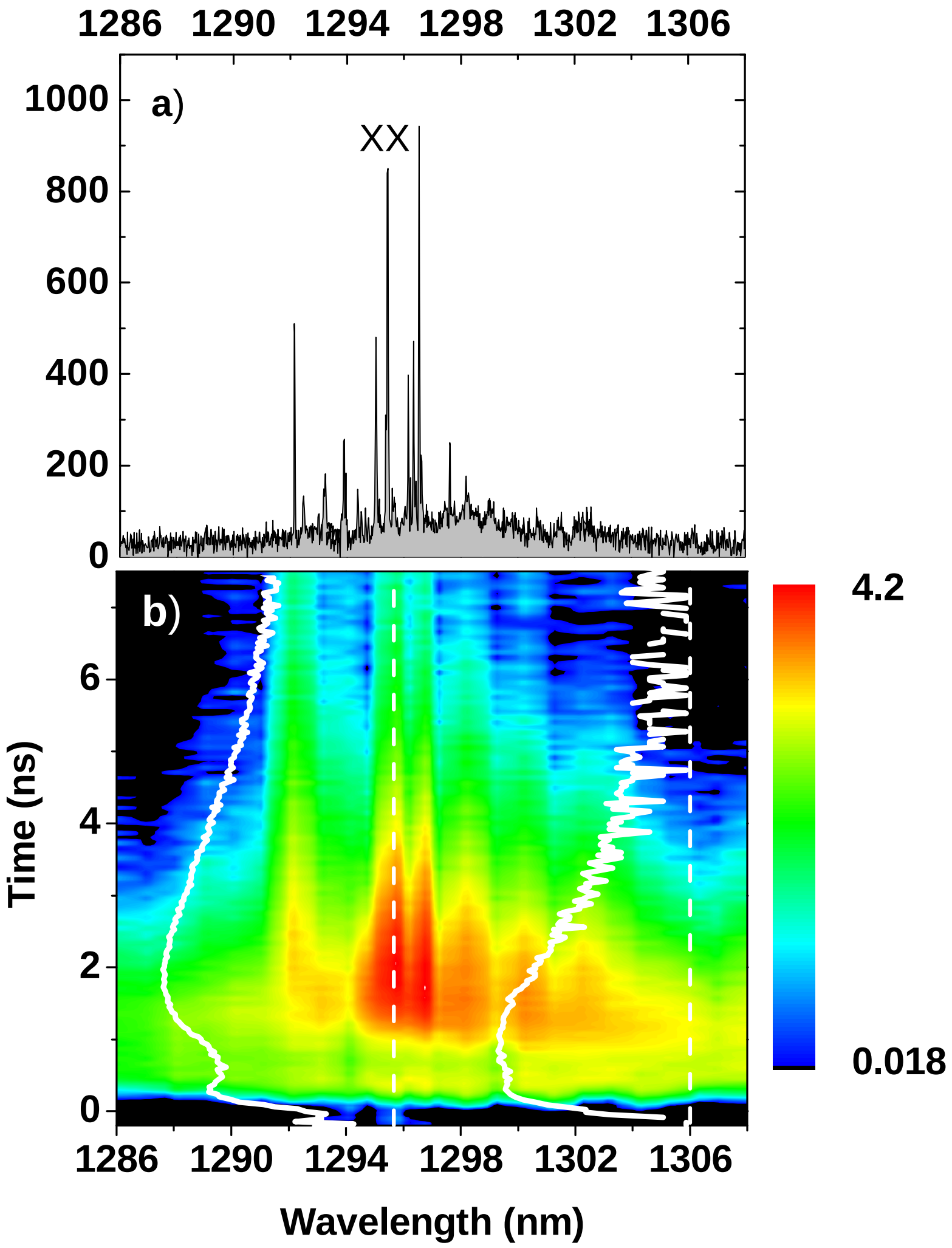

(Color online) Figure 2: (a) Spectrum of a single QD and (b) time-resolved experiments performed on the same QD. Two time-resolved experiments are shown (white curve): measurement on the biexciton line at 1295.5 nm and background emission at 1306 nm.

In the first phase of the decay, (<1ns), a continuous emission and featureless is observed from 1286 to 1308 nm. Then, from 1 to 3 ns, the emission is still broad but the intensity of the emission is stronger around 1295 nm where the majority of the single lines are observed. After 3 ns, the background emission progressively disappears and, at the end, the QD emission comes mainly from the single lines of the QD. The decays at two different wavelengths, corresponding to a single line (1295.5 nm), attributed to a biexciton [20] and to the featureless background (1306 nm), are also shown in Figure 2 in a logarithmic scale. The biexciton emission is clearly delayed as compared with the background: the maximum of intensity is reached after 1.2 ns for the background emission and 2 ns for the biexciton. The delayed biexciton emission clearly shows that (multi)excitonic lines can only take place after recombination of carriers in the 2D WL continuum, which strongly supports our interpretation of the origin of the background emission. This behavior also agrees with the anticorrelation between excitonic line and cavity mode observed in Ref. 2. As shown on Figure 2, the background emission is a complex phenomenon, with a frequency-dependent decay dynamics. Indeed, the different energies within the broad emission are connected to different charge densities and configurations of the carriers surrounding the QD. The temporal decay of this carrier population leads to a reshaping of the background emission as a function of time, ultimately resulting in clean (multi)excitonic lines.

In further support of our interpretation, we aim at providing direct evidence that the background emission is related to the presence of carriers in the WL. To this aim, we investigate the effect of an electric field on the emission spectrum in a photonic crystal (PhC) diode structure under reverse bias. The diode consists of a 370-nm thick GaAs/AlGaAs heterostructure with p- and n-contact layers on the two sides, incorporating a single layer of low density InAs QDs. The fabrication process is described in Ref 21.

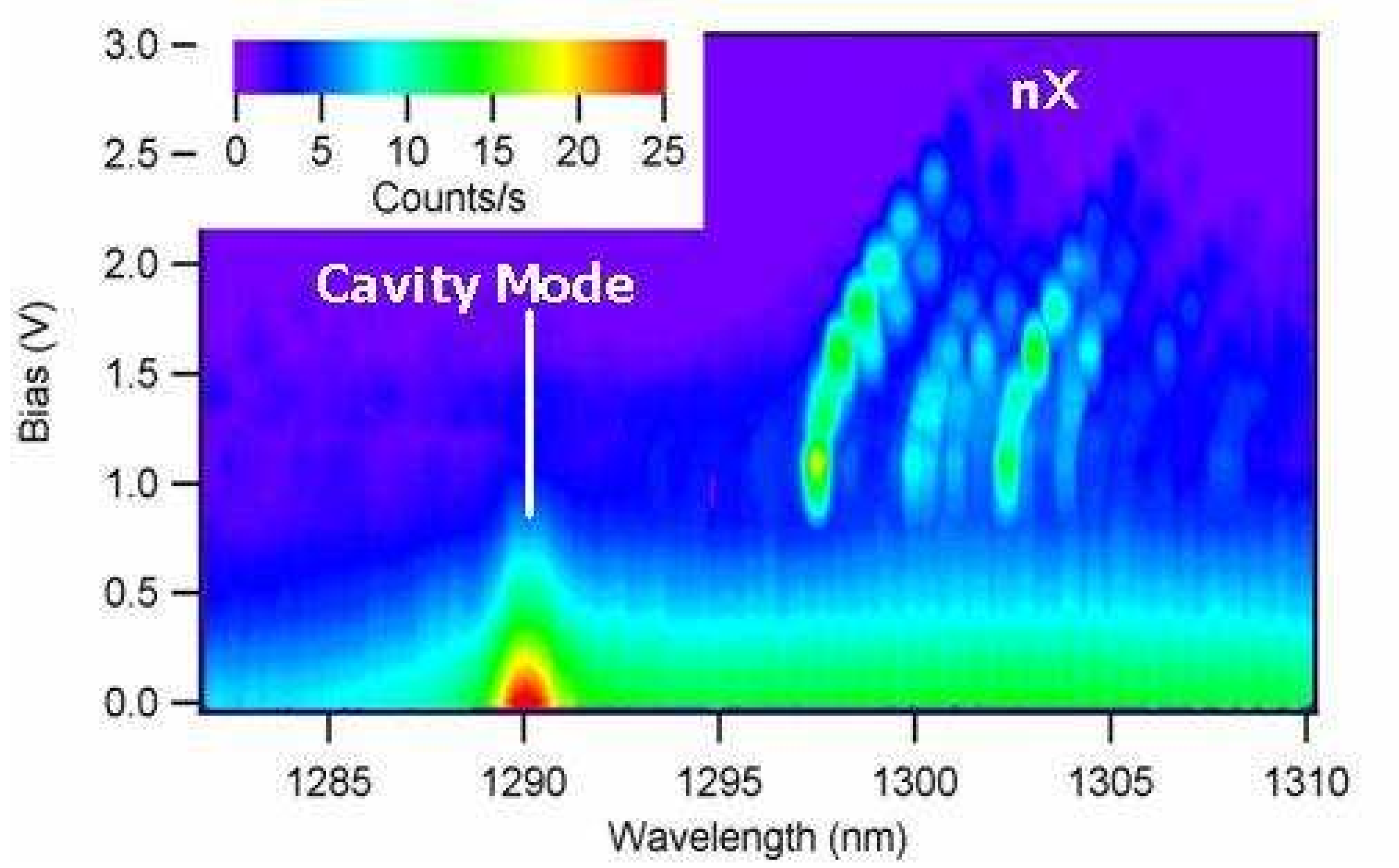

(Color online) Figure 3: Observation of the ground state emission of the PhC diode as a function of the bias voltage.

The photoluminescence of a L3 PhC diode has been studied using a cw 660 nm laser ($P_{av}$=5μW) under an applied voltage. PL spectra are presented in Fig. 3(b) as a function of the applied voltage $V_{bias}$ (defined as positive in reverse bias). For a small reverse bias (0-0.5V) a cavity mode is observed at 1290 nm with a quality factor of 850 along with a wide and unstructured background emission. A strong modification of the spectrum is observed when a bias voltage > 1V is applied: the cavity mode and the broad background disappear and (multi)excitonic lines ("nX") are observed. Our attribution of cavity and excitonic lines is confirmed by the field dependence of their energy. Indded, while the cavity energy does not vary with the voltage, the excitonic lines show a large Stark shift, corresponding to a dipole moment $p = 1.1 \times 10^{-28} Cm$ and a polarisability $\beta = -4 \times 10^{-36} Cm^2 V^{-1}$, close to values typically observed for InAs/GaAs QDs [22-24].

The intriguing observation of the disappearance of the cavity mode with applied field, and the simultaneous appearance of excitonic lines, clearly indicates that the cavity peak is associated to less confined, higher-energy carriers. These carriers are more easily swept away by the electric field than confined carriers in the QD, which results in the cleaning up of the spectrum and emergence of the excitonic lines. The observed electric-field dependence confirms our interpretation of the cavity emission and, at the same time, provides a means of controlling the QD charge environment and thus retrieving the ideal atom-cavity coupling.

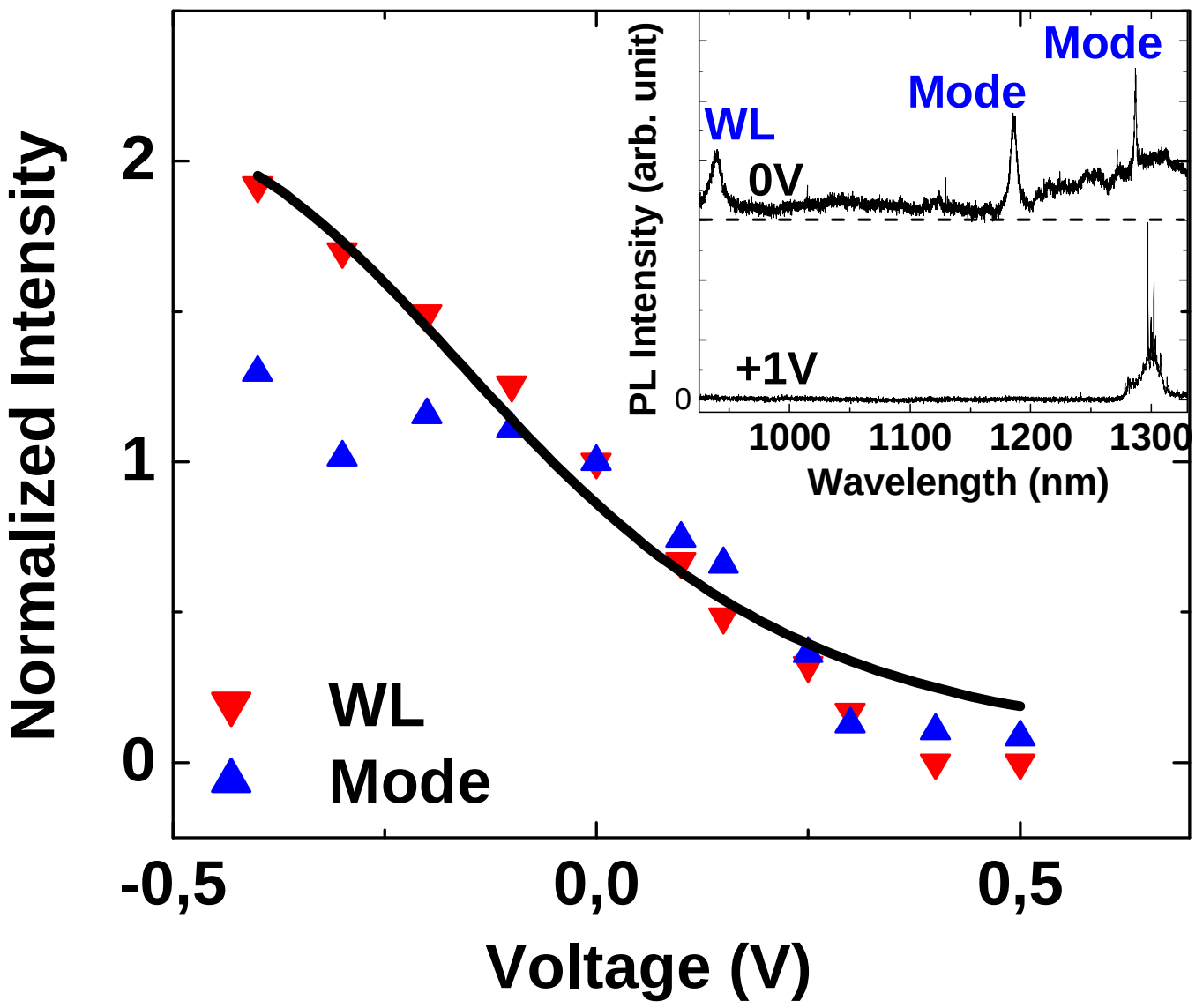

(Color online) Figure 4: Evolution of the wetting layer and the cavity mode as a function of the reversed bias. Inset: Emission of the PhC cavity for $V_b$=0 and $V_b$=1V.

In order to confirm that these additional carriers are indeed located in the WL, we have studied the WL and cavity emission as a function of the electric field. The inset in Figure 4 shows the micro-PL of the PhC diode over a wide spectral range for two different voltages. At 0V, two cavity modes are observed: the studied mode emitting at 1290 nm and a second one at 1185 nm. Additionally a strong emission of the wetting layer (WL) is observed at 940 nm consistent with previous observations on similar QDs [25]. In contrast, with a 1V applied bias both the WL and cavity peaks disappear, and single excitonic lines emerge. The integrated intensities of the wetting layer and of the cavity mode at 1.29 μm, reported in Fig 4, show the same dependence on the electric field confirming that they are both related to the same population. If we assume that the dynamics of the carriers in the wetting layer is driven by a carrier lifetime $\tau_0$ (including all field-independent radiative, non-radiative and capture processes) and a tunneling channel with a, field-dependent time constant $\tau_T(F)$ , the intensity of the WL peak as a function of the electric field F is given by $I(F) = I_0 / (1 + \tau_0 / \tau_T(F))$. In our case, the tunneling channel $\tau_T(F)$ is the tunneling rate through a triangular barrier (inset of Fig. 4(b)) which equals to [26]:

$$\tau_T^{-1}(F) = \frac{eF}{4\sqrt{2m^* V_{barrier}}} \exp\left( -\frac{4\sqrt{2m^*}}{3e\hbar F} V_{barrier}{}^{3/2} \right)$$

The tunneling rate is mainly related to the escape of electrons, due to their smaller effective mass as compared to the heavy holes. The experimental results are fitted using a single fitting parameter $V_b$, and fixing $m^* = 0.063 m_0$ (with $m_0$ the electron mass) and $\tau_0 = 400 ps$ [25], providing a value $V_b = 135 meV$. The energy gap discontinuity between the wetting layer and the bulk GaAs being equal to 200 meV, we can conclude that $V_{barrier}$ is the energy spacing between the WL electron ground state and the GaAs conduction band edge.

Due to the multiexcitonic nature of the background emission, a cavity mode pumped by the background should behave like a classical emitter. The auto correlation experiments performed on cavity modes indeed reveal a classical emission [2] under off-resonance excitation conditions similar to those used in our experiments. In contrast, single-photon emission from the cavity peak has been observed for quasi resonant excitation [5,9,11]. In the latter case, few carriers are created in the WL and the cavity mode is only pumped by the dephasing processes of a single exciton line, resulting in a non classical statistics.

In conclusion, the investigation of the dynamics of a QD-cavity system, and of its dependence on the electric field, has provided strong evidence that the a non-resonance cavity mode can be pumped by a broad QD emission originating from the Coulomb interaction between confined QD excitons and free carriers in the wetting layer. The application of an electric field removes the WL carriers, and therefore brings the QD closer to an ideal two-level system. This understanding and control of the charge environment of the QD is a key tool for the implementation of quantum information processing protocols based on the QD-cavity system. This mechanism for nonresonant coupling, observed here for relatively large detuning, does not exclude the existence of other phenomena such as dephasing processes already observed for small detuning. The relative influence of the different coupling processes depends on the detuning, on the excitation conditions and on the sample temperature.

**Acknowledgements**
We thank Dr. L. Lunghi (CNR) for nanopatterning of the gold apertures. We acknowledge funding from the EU-FP6 IP “QAP” Contract No. 15848, the Swiss

National Science Foundation through the "Professeur boursier" program, the Network of Excellence "ePIXnet," and the Italian MIUR-FIRB program.